\documentclass{article}

\usepackage{PRIMEarxiv}

\usepackage[utf8]{inputenc} 
\usepackage[T1]{fontenc}    
\usepackage{hyperref}       
\usepackage{amsmath}
\usepackage{url}            
\usepackage{booktabs}       
\usepackage{amsfonts}       
\usepackage{nicefrac}       
\usepackage{microtype}      
\usepackage{lipsum}
\usepackage{fancyhdr}       
\usepackage{graphicx}       
\usepackage{algorithm}
\usepackage{algorithmic}

\usepackage{color}
\usepackage{amsmath}
\usepackage{multirow}
\usepackage{array}
\usepackage{tablefootnote}
\usepackage{cleveref}
\graphicspath{{media/}}     

\title{
M6(GPT)3: Generating Multitrack Modifiable Multi-Minute MIDI Music from Text using Genetic algorithms, Probabilistic methods and GPT Models in any Progression and Time signature
\thanks{\textit{\underline{Citation}}: 
\textbf{J. Poćwiardowski, M. Modrzejewski and M. S. Tatara, "M6(GPT)3: Generating Multitrack Modifiable Multi-Minute MIDI Music from Text using Genetic Algorithms, Probabilistic Methods and GPT Models in any Progression and Time Signature", 2025 IEEE International Conference on Multimedia and Expo Workshops (ICMEW), Nantes, France, 2025, Pages 1-6 DOI:10.1109/ICMEW68306.2025.11152218.}} 
}

\author{
  Jakub Po\'cwiardowski, Mateusz Modrzejewski \\
  Institute of Computer Science \\
  Warsaw University of Technology \\
  Warsaw\\
  \texttt{jakub.pocwiardowski.stud@pw.edu.pl, mateusz.modrzejewski@pw.edu.pl} \\
   \And
  Marek S. Tatara \\
  Department of Robotics and Decision Systems \\
  Gdansk University of Technology \\
  Gdansk\\
  \texttt{marek.tatara@pg.edu.pl} \\
}

\begin{document}
\maketitle

\begin{abstract}
\begin{quote}
This work introduces the M6(GPT)3 composer system, capable of generating complete, multi-minute musical compositions with complex structures in any time signature, in the MIDI domain from input descriptions in natural language. The system utilizes an autoregressive transformer language model to map natural language prompts to composition parameters in JSON format. The defined structure includes time signature, scales, chord progressions, and valence-arousal values, from which accompaniment, melody, bass, motif, and percussion tracks are created. We propose a genetic algorithm for the generation of melodic elements. The algorithm incorporates mutations with musical significance and a fitness function based on normal distribution and predefined musical feature values. The values adaptively evolve, influenced by emotional parameters and distinct playing styles. The system for generating percussion in any time signature utilises probabilistic methods, including Markov chains. Through both human and objective evaluations, we demonstrate that our music generation approach outperforms baselines on specific, musically meaningful metrics, offering a viable alternative to purely neural network-based systems.
\end{quote}
\end{abstract}

\section{Introduction}

Musicians around the world are constantly searching for inspiration to create groundbreaking and unique music. There has been a lot of work in recent years concerning generating music with the usage of AI techniques, both in symbol and audio domain. The latest state-of-the-art methods for generating music from text focus on the audio domain with works such as OpenAI's Jukebox \cite{dhariwal2020jukebox}, Google's MusicLM \cite{agostinelli2023musiclm} or Meta's MusicGen \cite{copet2023simple}. For symbolic composition, there are not many works capable of generating music from text, as there is not much text-symbolic music data and symbolic music is much more difficult to describe in natural language, especially for non-musicians. Some of the works trying to accomplish that are MuseCoCo \cite{lu2023musecoco} and BUTTER \cite{zhang2020butter}. The most advanced symbolic models without prompt conditioning involve usage of Transformers in works such as \cite{huang2018music, huang2020pop}, which are also used to compose multitrack music \cite{ens2020mmm, dong2023multitrack}.

All the current state-of-the-art methods use neural networks to effectively replicate human music composition abilities. However, they depend on broad datasets, leading to reliance on dominant structures like 4/4 meter and common chord progressions. Furthermore, text-to-music models often fail to respond to specific musical terms, and audio-based music generation models produce outputs that are hard to edit further. To address these limitations, we introduce the M6(GPT)3 system, designed for flexible multi-track song generation and editing in symbolic format.


Our contributions may be summarized as follows:
\begin{itemize}
    \item We propose M6(GPT)3, a novel solution for generating and editing full multi-track songs in symbolic format without relying on dominant musical structures. We integrate genetic algorithms and probabilistic methods to generate MIDI tracks for specific song sections.
    \item We evaluate M6(GPT)3 using subjective and objective, musically meaningful metrics.
    \item We release demo examples and Python code to facilitate reproducibility and further research.
\end{itemize}

\section{Related Work}

\subsection{Large language models for music generation}
The recent revolution of Large Language Models (LLMs) invoked experiments concerning their abilities in tasks they were not initially created for. Examples of LLM usage for music composition include ComposerX \cite{deng2024composerx} which uses a multi-agent approach and chain-of-thought, where distinct instances of GPT-4 take care of interpreting user inputs, creating melodies, harmonising, and choosing instruments.

\subsection{Genetic algorithms for music generation}

Genetic or evolutionary algorithms optimize melodies based on specified criteria, such as user feedback on quality. An early, well-known example of this approach is GenJam \cite{biles1994genjam} meant for generating jazz solos. Due to the time-consuming nature of human feedback, latter approaches focus on developing objective criteria for fitness functions. \cite{towsey2001towards} proposes 21 qualities related to melody, pitch, tonality, rhythm, and repetitiveness for this purpose. \cite{liu2013evolutionary} uses music theory and charts, while \cite{kowalczuk2017evolutionary} employs Gaussian distribution to model the criteria of the cost function.

\subsection{Conditioning music generation}

Non-neural music generation conditioning is usually based on predefined rules and emotions. The emotions are usually represented either as discrete emotions or as a point on a valence-arousal plane. \cite{wallis2011rule} presents a rule-based mapping of valence and arousal values on musical qualities such as beats-per-minute (BPM), scales or pitch range. On the other hand, \cite{kuo2015development} proposes rules and fuzzy logic to map 12 emotions on tempo and other qualities. The predefined decision tree for chord selection is based on the circle of fifths, while the melody is created for the chords using genetic algorithms.

\section{Structure of M6(GPT)3}

The architecture was planned in order to have a separate components, which can be tested and developed independently, yet together create an innovative system. The system's pipeline involves the following steps: 
\begin{enumerate}
    \item \textbf{Predicting} composition's structure and parameters from text using LLM,
    \item \textbf{Generating} melodic and drum tracks based on the structure and parameters provided by the LLM,
    \item \textbf{Integrating} generated tracks into a MIDI file.
\end{enumerate}

\begin{figure*}[ht]
\centering
\includegraphics[width=0.85\textwidth]{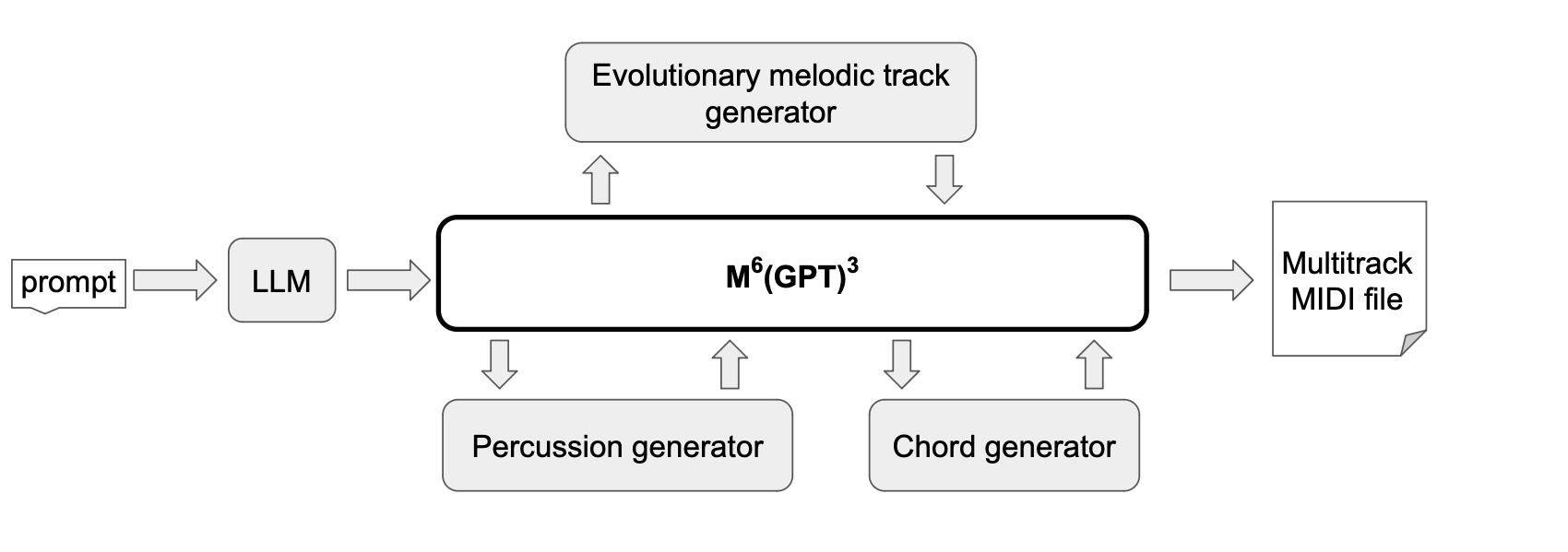} 
\caption{The system architecture illustrating all key components. The MIDI Generator receives composition parameters, which it forwards to individual track generators. These generators produce tracks (each of them independently, based on the same input describing song structure) that are combined and encoded into a single MIDI file.}
\label{fig:diagram}
\end{figure*}

The system structure is shown in \cref{fig:diagram}. While the LLM receives no direct feedback, the generated output and prior prompts provide context for further modification.

\subsection{Predicting structure from text}

To generate song structure and parameters, an LLM is used. We chose the GPT family because of its promising music theory knowledge and ability to use it via openai API, making it accessible from any computer, contrary to locally hosted LLMs requiring high computing resources. Note that the system's architecture features an interchangeable LLM component, enabling the integration of more advanced models.

To obtain the structure in an appropriate form, the model is provided with precise instructions as a system prompt:
\begin{quotation}
\itshape
"You are a music composing system. The user will ask you about the song they want to generate and your task is to respond with output of JSON file and JSON file only (...)"
\end{quotation}


The complete JSON template contains:

\begin{itemize}
    \item \textbf{Song name},
    \item \textbf{Song sections} with details on BPM, time signatures, scales, tracks (instrument and technique), chord progressions, and repeats,
    \item \textbf{Section placement} in the song, including valence-arousal points\footnote{Valence: $\langle-1,1\rangle$, Arousal: $\langle0,1\rangle$, chosen for LLM comprehension.},
    \item \textbf{Composer's note} explaining artistic choices.
\end{itemize}

The model is instructed about the available set of scales, instruments, and chord types with precise indications to stick to this set. Another instruction is to maintain a coherent structure of the whole song so that the instrumentation and style of consecutive sections do not vary too much from each other unless prompted to do so.

The language model allows for iterative composition and editing based on user prompts. Users specify changes, and the model will adjust the structure accordingly, whether altering exact specified parameters or adapting the parameters to broader descriptions, such as sophistication, mood, or specific composer's style.


\section{Evolutionary Melodic Track Generator}

The melodic sections in the presented multitrack systems are generated with genetic algorithms. Using this method, three types of tracks are generated:
\begin{enumerate}
    \item \textbf{Melodies}
    - the most unrestricted tracks, mostly mid-range but capable of high and low notes.
    \item \textbf{Bass}
    - playing in low frequencies, usually more restricted and more repetitive than melodies.
    \item \textbf{Motif}
    - usually playing in high frequencies, being the most repetitive track, adding texture to the sound
\end{enumerate}

\subsection{Encoding and Genetic Operations}

To encode musical notes for optimization with a genetic algorithm, we used the same approach as \cite{kowalczuk2017evolutionary,liu2013evolutionary}. Notes are represented as MIDI values from 0 to 127, with rests marked as -1 and note extensions as -2. This encoding allows for different note durations while preserving a consistent bar length, which is essential for a multitrack system with multiple sections.

The operations used in our genetic algorithm are:
\begin{itemize}
    \item Random initialization,
    \item One-point crossover,
    \item Tournament selection,
    \item Mutations with musical significance,
    \item Fitness function based on the normal distribution.
\end{itemize}

\subsection{Mutations with Musical Significance}

We use mutations with musical significance to introduce musical features into melodies, which by default consist only of notes with uniform, minimum length. The choice of musical mutations was inspired mainly by \cite{kowalczuk2017evolutionary, biles1994genjam}. Table \ref{tab:mutations-detailed} shows exemplary mutations used in our systems with examples of their impact on mutated melody.

\begin{table*}[h!]
\small
\centering
\resizebox{\linewidth}{!}{
\begin{tabular}{|l|p{8.5cm}|c|c|c|c|c|c|c|c|}
\hline
\textbf{Mutation} & \textbf{Effect} & \multicolumn{8}{c|}{\textbf{Example sequence}} \\ \hline
Original           & - & 81 & 58 & 46 & 58 & 46 & -2 & -2 & 61 \\ \hline
Interval           & Creates a random interval of -12 to 12 semitones between two notes. & 81 & \textbf{89} & 46 & 58 & 46 & -2 & -2 & 61 \\ \hline
Transpose          & Shifts a sequence segment by a random interval in $\langle-12,12\rangle$ range. & 81 & 58 & 46 & \textbf{61} & \textbf{49} & \textbf{-2} & \textbf{-2} & \textbf{64} \\ \hline
Extend             & Extends a random note, shortening (or removing) the preceding. & 81 & 58 & 46 & 58 & 46 & -2 & \textbf{61} & \textbf{-2} \\ \hline
Rest              & Converts random note to a rest or a rest to a note. & 81 & 58 & 46 & 58 & \textbf{-1} & -2 & -2 & 61 \\ \hline
Long note          & Changes a random series of notes into extensions (-2). & 81 & 58 & \textbf{-2} & \textbf{-2} & 46 & -2 & -2 & 61 \\ \hline
-2 to note         & Converts a random extensions (-2) into a random note. & 81 & 58 & 46 & 58 & 46 & -2 & \textbf{41} & 61 \\ \hline
Length norm.       & Divides long notes in two or extends short notes\tablefootnote{We treat a quarter note as the "normal" length. Notes longer than this are randomly split, while shorter ones are extended.}. & 81 & 58 & 46 & 58 & \textbf{46} & \textbf{46} & -2 & 61 \\ \hline
Sort               & Sorts a random slice of a sequence in a random direction. & \textbf{46} & \textbf{46} & \textbf{58} & \textbf{58} & \textbf{81} & -2 & -2 & 61 \\ \hline
Repeat 1           & Randomly chooses a sequence and pastes it to another measure. & \emph{81} & \emph{58} & 46 & 58 & \textbf{81} & \textbf{58} & -2 & 61 \\ \hline
Repeat 2           & Randomly chooses a sequence and pastes it immediately after. & \emph{81} & \emph{58} & \textbf{81} & \textbf{58} & 46 & -2 & -2 & 61 \\ \hline
\end{tabular}
}
\caption{Mutation with musical significance (here, to illustrate the mutations in a table, we treat a measure as a 4-element sequence instead of factual 16).}
\label{tab:mutations-detailed}
\end{table*}

\subsection{Fitness Function}

When designing the fitness function, we adapted the approach from \cite{kowalczuk2017evolutionary}, which employs a Gaussian distribution. In our implementation, the distribution's mean represents the target metric value. The melody's fitness score is determined by the weighted sum of each metric's distribution value at the point corresponding to the actual measured metric value of the melody. The metrics used for genetic algorithm optimization include: the mean percentage of unique notes and intervals per measure; percentages of dissonant intervals, intervals larger than an octave, notes in the scale, notes in the current chord and rests in a sequence; tonal range; mean percentage of unique note lengths per measure; average note pitch and its deviation; length of strong beat notes; melodic contour; off-beat notes; normalized average interval size; logarithm of average note length and its deviation; counts of consecutive intervals within (1,3) and short notes (eighth or shorter); length of repeated fragments; and the number of measures beginning with a root note. All measured values are normalized to the (0, 1) range. The formula for fitness function calculation is presented in \cref{eq:main_equation}.


\begin{equation} \label{eq:main_equation}
f(x) = \sum_{i=1}^{n} w_i \exp\left( -\frac{(r_i(x) - \mu_i)^2}{2\sigma_i^2} \right)
\end{equation}
where:
\begin{itemize}
\item \( x \) is the melodic sequence,
\item  \( n \) is the number of features being measured in the sequence,
\item \( w_i \) is the weight for $i$-th feature,
\item \( \sigma_i \) is the standard deviation value for $i$-th feature,
\item \( \mu_i \) is the desired value for $i$-th feature,
\item \( r_i \) is the measured value of $i$-th feature in the sequence.
\end{itemize}



To ensure harmonic coherence in multitrack compositions, we introduce a metric to assess inter-track dissonance. Based on \cite{liu2012polyphonic}, our method accounts for rests and evaluates intervals modulo 12 using \cref{tab:harmony_intervals}. The total score is the average of these values, normalized via the $tanh$ function as in \cref{eq:normalization}.

\begin{table}[h]
\centering
\begin{tabular}{lc}
\hline
Interval (semitones) & Score \\
\hline
0, 3, 4, 8, 9 & 8 \\
5, 7 & 15 \\
1, 2, 10, 11 & -20 \\
6 & -30 \\
Rest in one of the two tracks & 10 \\
Rest in both tracks & 0 \\
\hline
\end{tabular}
\caption{Scores assigned to various intervals between tracks.}
\label{tab:harmony_intervals}
\end{table}

\begin{equation}
\label{eq:normalization}
\tanh\left(\frac{a_e}{10}\right).
\end{equation}

The scoring system yields a range from -30 to 15 before normalization. To ensure that the highest scores do not disproportionately influence the results, we used 10 as the denominator in the normalization function. This approach rewards a variety of pleasing interval combinations while modestly favoring the most harmonic ones. Conversely, dissonances incur a swift penalty.

\subsection{Melodic Tracks Modes}

To support the diverse characteristics of melodic sections and assist LLM decision-making, we arbitrarily define three melodic sections and 11 generation modes, which influence genetic algorithm parameters.

For the \textbf{main melody}, the system operates in two modes: 
\emph{Melody} (conservative tonal range and dynamics) and \emph{Solo} (greater freedom, featuring fast and short notes).

\textbf{Bass tracks} include four modes:
\emph{Short riff} (one-measure riff repeated and shifted to fit chord roots),
\emph{Long riff} (two-measure riff with similar behavior),
\emph{Bassline} (lowest chord roots with randomized transitions and mutations for emotion),
and \emph{Repetitive bassline} (bassline with same mutations per bar).

\textbf{Motifs} offer five modes:
\emph{Long motif} (one measure, repeated and shifted),
\emph{Opening motif} (half-measure with rests in the second half),
\emph{Closing motif} (rests in the first half),
\emph{Repeated motif} (half-measure repeated),
and \emph{Short repeated motif} (quarter-note length, repeated and trimmed).

These modes, combined with valence-arousal values, define the genetic algorithm parameters. The rules draw inspiration from works \cite{friberg2006pdm, miyamoto2020music, wallis2011rule} and our own experiences. Table \ref{tab:merged_metrics} presents the fitness function metrics, their desired values for specific tracks, and the influence of emotions.

\begin{table}[h]
\centering
\resizebox{\columnwidth}{!}{%
\begin{tabular}{|c|c|c|c|c|c|c|c|}
\hline
\textbf{Metric}              & \textbf{Melody} & \textbf{Solo} & \textbf{Bass}  & \textbf{Motif} & \textbf{EI}        & \textbf{V}       & \textbf{A}       \\
\hline
Mean percentage of unique notes per measure            & Low      & High  & Med  & High  & None                      & -           & -           \\
\hline
Mean percentage of unique intervals per measure        & Med      & High  & High & High  & Low                       & $\uparrow$  & $\uparrow$  \\
\hline
Percentage of dissonant intervals             & Low      & Low   & Low  & Low   & Med                       & $\downarrow$& -           \\
\hline
Percentage of intervals bigger than octave      & Zero     & Zero  & Zero & Zero  & None                      & -           & -           \\
\hline
Percentage of notes in scale          & High     & High  & High & High  & Low                       & -           & $\downarrow$\\
\hline
Percentage of notes in current chord         & Med      & Med   & High & High  & Low                       & $\downarrow$& $\downarrow$\\
\hline
Pitch range             & Low      & High  & Med  & Med   & Med                       & -           & $\uparrow$  \\
\hline
Percentage of rests in a sequence                  & Low      & Low   & -    & -     & Med                       & $\downarrow$& $\downarrow$\\
\hline
Mean percentage of unique note lengths per measure     & Low      & Med   & -    & -     & None                      & -           & -           \\
\hline
Average note pitch              & Med      & Med   & Low  & Med   & Med                       & $\uparrow$  & $\uparrow$  \\
\hline
Pitch deviation         & Med      & Med   & Med  & -     & Med                       & -           & $\uparrow$  \\
\hline
Length of strong beat notes       & High     & High  & High & -     & Med                       & -           & $\downarrow$\\
\hline
Melodic contour         & Med      & Med   & Med  & Med   & High                      & $\uparrow$  & -           \\
\hline
Off-beat notes          & Low      & Low   & Low  & -     & None                      & -           & -           \\
\hline
Average interval size (normalized to one octave)      & Med      & Med   & Med  & -     & Med                       & $\uparrow$  & -           \\
\hline
Logarithm of average note length        & Med      & Med   & Med  & -     & Med                       & -           & $\downarrow$\\
\hline
Deviation of logarithms of note lengths        & Med      & Low   & Low  & -     & Med                       & $\downarrow$& $\uparrow$  \\
\hline
Amount of consecutive intervals in range (1,3)   & High     & High  & -    & Med   & None                      & -           & -           \\
\hline
Amount of consecutive short notes (eight or shorter)        & Med      & High  & -    & -     & Med                       & -           & $\uparrow$  \\
\hline
Length of repeated fragments      & Med      & Med   & Med  & -     & Low                       & -           & $\uparrow$  \\
\hline
Measures beginning with a root note              & -        & -     & Med  & Med   & Low                       & -           & $\downarrow$\\
\hline
\end{tabular}
}
\caption{Desired metric values for melody, solo, bass, and motif, with emotion impact (EI) on their values. Arrows in V (valence) and A (arousal) show direction of change (proportional or inversely proportional).}
\label{tab:merged_metrics}
\end{table}

\section{Percussion Generator}

To generate songs in odd time signatures, Deep Learning models like Transformers may not be ideal, as training datasets are heavily biased toward 4/4. Our analysis of the well-known Groove dataset \cite{groove2019} revealed that 98.96\% of its drum loops are in 4/4 time. Given the relative simplicity and repetitiveness of drum patterns in many songs, we opted for a rule-based system incorporating probability and Markov Chains.

\subsection{Data Representation}

To encode drum states at a given time, we employed a binary representation of drum components proposed in \cite{choi2016text}. Here, we use 12 components of percussion, namely: closed hi-hat, open hi-hat, bass drum, snare, 5 toms, crash, ride, and bell. To present an example 100000000000 and 001000000000 represent the closed hi-hat and bass drum respectively, while 101000000000 represents playing these components simultaneously. In default, one state corresponds to a sixteenth note.

\subsection{Drum Patterns}

We analyzed fundamental drum patterns from various sources to construct probability tables for bass drum and snare placement in time signatures from 2/4 to 9/8. For unsupported signatures, we use recursive decomposition (e.g., 13/8 splits into 7/8 and 6/8), while shorter-note signatures (e.g., 11/16) inherit patterns from longer equivalents (e.g., 11/8). Each drum state is encoded in sixteenth-note intervals, requiring silent padding: three silent states per beat for quarter-note-based signatures, one for eighth-note-based signatures, and so forth.

Hi-hats emphasize tempo, appearing on quarter, eighth, or sixteenth notes, with patterns shaped by the section's emotion. Open hi-hats often replace the final closed hi-hat in a measure to mark transitions, influenced by the arousal parameter. Ride cymbals and bells typically follow quarter-note pulses, offset by an eighth, with emotional variations. Crash cymbals primarily occur at measure starts, rarely appearing elsewhere.




\subsection{Toms and Drum Fills}

Drum fills are generated using Markov chains, guided by fill probability and length. Each state represents a 5-bit pattern for tom activations, with a separate state for the snare.

After assembling the 12-bit drum sequences, we enhance diversity and realism by modifying patterns. Bass and snare hits may repeat probabilistically. If hand-played elements (bits 1, 2, 5–12) exceed two in a state, random bits are zeroed to reduce the count. Additionally, bits 5–12 are randomly zeroed per state with a set probability.



\subsection{Percussion Modes}



To enable the drum system to be effectively driven by the LLM, we define three playing modes: \emph{Only Beat} (no drum fills), \textit{Drum Solo} (only drum fills), and \emph{Standard} (drum fills depend on emotional values, with higher probability in the final measure of a section).

Additionally, we offer three drum kits using General MIDI sounds: a \emph{standard kit}, an \emph{ethnic kit}, and an \emph{orchestral kit}. The generation process is identical across kits, but binary indices correspond to different components based on the selected kit.

\section{Chord generator}



Chord progressions form the composition's foundation, defined per section by the language model. The chord track is generated in one of three modes: Continuous (sustained chords), Repeated (notes played with arousal-dependent lengths, omitting up to 40\% of repetitions), or Arpeggio (notes played sequentially with durations from a sixteenth to a half note based on arousal).

Since the LLM does not specify octaves or voicing, we adjust inversions and note counts using valence-arousal values, partially following \cite{wallis2011rule}. Chord size varies from two notes (low arousal) to six (high arousal), with adjustments: removing the fifth for low arousal or large chords, adding an octave-above note for high arousal if the chord is small, and including both an octave and fifth at extreme arousal levels.

For octave placement, we modify \cite{wallis2011rule}’s approach to fit our multitrack system, shifting the note concentration range from A2 to A4 to avoid overlap with the melody. The center of gravity is determined by valence, and chords are optimized to maintain spacing and minimize overlap. Conditioning via valence and arousal is detailed in \cref{tab:state_vs_condition}.

\begin{table}[h!]
\centering
\resizebox{0.7\linewidth}{!}{
\begin{tabular}{|c|c||c|c|}
\hline
\textbf{Valence State} & \textbf{Pitch Register} & \textbf{Arousal State} & \textbf{Voicing Size} \\ 
\hline
Min & Centered on A2  & Min & 2 Notes \\
Max & Centered on A4 & Max & 6 Notes \\
\hline
\end{tabular}
}
\caption{Influence of Emotional Values on Chord Parameters.}
\label{tab:state_vs_condition}
\end{table}

\section{Experiemental evaluation}

\subsection{Experiemental Setup}

In our experiments, we employ the gpt-4-1106-preview model with a temperature of 0. For all generations we use population size of 256 with 100 generations, tournament size of 4, 0.3 mutation rate and 0.9 crossover rate.
\subsection{Subjective Listening Test}
To assess the quality of the music generated by our model, we conducted a listening test with 23 participants, who were recruited from people who have basic musical knowledge and understanding of basic musical concepts.

The listening test was divided into two segments. In the first segment (\cref{tab:subjective-results}), participants evaluated 15 compositions from 3 different systems. Each composition was rated on a 1-to-5 scale across five criteria: 1) richness and diversity, 2) memorability, 3) entertainment value, 4) emotional conveyance, and 5) inspirational quality. The comparison systems were ComposerX \cite{deng2024composerx} and MMT \cite{dong2023multitrack}. We selected 5 "best examples" that were more than a minute in length from the websites of both systems, randomly mixed them with our model's 5 best compositions, and presented them to participants for evaluation.

\begin{table*}[th!]
\centering
\resizebox{\linewidth}{!}{
\begin{tabular}{|c|c|c|c|c|c|c|}
\hline
& \textbf{RD} & \textbf{M} & \textbf{EV} & \textbf{EC} & \textbf{I} & \textbf{Avg} \\
\hline
$\text{M}^\text{6}(\text{GPT})^\text{3}$ (ours) 
& \textbf{3.40 ± 0.18 }
& \textbf{3.05 ± 0.20 }
& \textbf{3.34 ± 0.20 }
& 3.31 ± 0.18 
& \textbf{3.36 ± 0.21 }
& \textbf{3.29 ± 0.09} \\
\hline
MMT \cite{dong2023multitrack} 
& 3.35 ± 0.19 
& 2.98 ± 0.20 
& 3.03 ± 0.22 
& \textbf{3.34 ± 0.19 }
& 3.30 ± 0.22 
& 3.20 ± 0.09 \\
\hline
ComposerX \cite{deng2024composerx} 
& 3.07 ± 0.20 
& 2.90 ± 0.21 
& 3.04 ± 0.22 
& 2.94 ± 0.23 
& 2.87 ± 0.25 
& 2.97 ± 0.10 \\
\hline
\end{tabular}
}
\caption{Comparison against the baseline models (\textbf{RD}: Richness and diversity, \textbf{M}: Memorability, \textbf{EV}: Entertainment value, \textbf{CE}: Emotional conveyance, \textbf{I}: Inspirational quality, \textbf{Avg}: Average of all criteria). Mean values and 95\% confidence intervals are reported. According to the listeners, M6(GPT)3 outperformed the other evaluated models in 4 out of 5 criteria, with its most significant advantage being the entertainment value of its compositions.}
\label{tab:subjective-results}
\end{table*}

\begin{table*}[th!]
\small
\centering

\resizebox{\linewidth}{!}{
\begin{tabular}{|c|c|c|c|c|c|}
\hline
 & \textbf{Tempo} & \textbf{Instruments} & \textbf{Mood} & \textbf{Structure} & \textbf{Overall} \\
\hline
M6(GPT)3 generation 
& 4.06 ± 0.28 
& 4.25 ± 0.24 
& 3.58 ± 0.32 
& 3.61 ± 0.41 
& 3.80 ± 0.25 \\
\hline
Actual described song from MIDICaps 
& \textbf{4.14 ± 0.21 }
& \textbf{4.38 ± 0.17} 
& \textbf{3.79 ± 0.23 }
& \textbf{4.09 ± 0.26 }
& \textbf{4.06 ± 0.18} \\
\hline
\end{tabular}
}
\caption{Comparison of original song from MIDICaps dataset and M6(GPT)3 generation from the same description evaluating how well each matches that description. Mean values and 95\% confidence intervals are reported. While the generations produced by M6(GPT)3 generally underperform compared to the actual described songs, their results demonstrate a notable proximity to the original compositions.}
\label{tab:subjective-results-midicaps}
\end{table*}

In the second test (\cref{tab:subjective-results-midicaps}), we compared our system's outputs to MIDICaps \cite{melechovsky2024midicaps}, the only large-scale public MIDI dataset with text descriptions. We randomly selected 5 full-length pieces from the dataset and generated 5 corresponding full-length songs using our system based on the same descriptions.
\footnote{Generated pieces can be heard on https://jpocwiar.github.io/M6-GPT3-Composer-Demo/.}
The participants were then surveyed to determine whether both the dataset compositions and the M6(GPT)3 generations accurately matched their descriptions regarding tempo, instruments, mood, structure, and overall impression.

\subsection{Objective Evaluation}

To further support the listening test, we follow \cite{dong2023multitrack} and evaluate pitch class entropy, scale consistency, and groove consistency using the MusPy framework \cite{dong2020muspy, dong2018musegan}. For our system, we assess these metrics in three different scenarios. The ``Prompt-driven'' scenario involves generating full songs from text using an LLM. The ``Focused''  scenario standardizes the structure with a 4/4 time signature, 120 bpm tempo, C-F-Am-F chord progression in C Major and middle emotional values across all generations. In the ``Randomized'' scenario, all parameters are random except for the chords, which belong to the same scale. We compare our results with those presented in \cite{dong2023multitrack} and display them in \cref{tab:objective-results}. We find that the outputs of $\text{M}^\text{6}(\text{GPT})^\text{3}$ have comparable musical traits to a reference set of real music.

\begin{table*}[th!]
\small
\centering
\resizebox{\linewidth}{!}{
\begin{tabular}{|c|c|c|c|}
\hline
 & Pitch Class Entropy & Scale Consistency (\%) & Groove Consistency (\%) \\
\hline
Ground truth (reference music)
& $2.974 \pm 0.018$ 
& $92.26 \pm 1.25$ 
& $93.05 \pm 1.00$ \\
\hline
MMM \cite{ens2020mmm} 
& $2.884 \pm 0.023$ 
& $93.13 \pm 0.49$ 
& $91.90 \pm 0.64$ \\
REMI+ \cite{von2022figaro} 
& $2.897 \pm 0.019$ 
& $93.12 \pm 0.51$ 
& $\mathbf{92.90 \pm 0.49}$ \\
MMT \cite{dong2023multitrack} 
& $2.802 \pm 0.025$ 
& $94.74 \pm 0.42$ 
& $92.09 \pm 0.49$ \\
\hline
$\text{M}^\text{6}(\text{GPT})^\text{3}$ - Prompt-driven (ours) 
& $\mathbf{2.907} \pm 0.067$ 
& $\mathbf{92.640 \pm 2.11}$ 
& $98.919 \pm 0.15$ \\
$\text{M}^\text{6}(\text{GPT})^\text{3}$ - Focused (ours) 
& $2.849 \pm 0.010$ 
& $98.804 \pm 0.14$ 
& $98.558 \pm 0.01$ \\
$\text{M}^\text{6}(\text{GPT})^\text{3}$ - Randomized (ours) 
& $2.886 \pm 0.021$ 
& $95.113 \pm 0.94$ 
& $98.890 \pm 0.11$ \\
\hline
\end{tabular}
}
\caption{Comparison with the baseline models with objective metrics following \cite{dong2023multitrack}. Mean values and 95\% confidence intervals are reported, with the closest values to the ground truth highlighted in bold. Across three established metrics, M6(GPT)3 with LLM-conditioned song structure achieves the best performance in two out of three cases.}
\label{tab:objective-results}
\end{table*}

\section{Conclusions}

In this work, we introduce M6(GPT)3, a novel method for generating music from text without relying on extensive music datasets. Both objective metrics and subjective studies show that this approach can produce realistic and inspiring compositions. To facilitate further research, we publish code and generated examples. We believe that as fully neural-based systems continue to evolve, there is an increasing need to emphasize optimization-based techniques to create innovative and creative tools. The combination of controllability and structural flexibility of our approach with neural advances such as style morphing has the potential to push the boundaries of music generation in the future.

\bibliographystyle{unsrt}  

\end{document}